# Modeling of NO sensitization of IC engines surrogate fuels auto-ignition and combustion

**J. Anderlohr [1,2], A.Pires da Cruz [1]**
**R. Bounaceur [2]. F. Battin Leclerc [2]**

[1]IFP, 1 et 4, Av. Bois Préau, 92852 Rueil Malmaison Cedex, France

[2]*Département de Chimie-Physique des Réactions, CNRS-INPL*
*ENSIC, 1, rue Grandville, BP 20451, 54001 Nancy Cedex, France*

## 1 Introduction

This paper presents a new chemical kinetic model developed for the simulation of auto-ignition and combustion of engine surrogate fuel mixtures sensitized by the presence of $NO_x$. The chemical mechanism is based on the PRF auto-ignition model (n-heptane/iso-octane) of Buda et al. [1] and the NO/n-butane/n-pentane model of Glaude et al. [2]. The later mechanism has been taken as a reference for the reactions of $NO_x$ with larger alcanes (n-heptane, iso-octane). A coherent two components engine fuel surrogate mechanism has been generated which accounts for the influence of $NO_x$ on auto-ignition. The mechanism has been validated for temperatures between 700 K and 1100 K and pressures between 1 and 10 atm covering the temperature and pressure ranges characteristic of engine post-oxidation thermodynamic conditions. Experiments used for validation include jet stirred reactor conditions for species evolution as a function of temperature, as well as diesel HCCI engine experiments for auto-ignition delay time measurements.

During the last decades, the European Union legislation has imposed more and more stringent restrictions on vehicle exhaust emissions. For this reason, pollutant emissions control is of primary role for internal combustion engine research and development. **S**econdary **A**ir **I**njection (SAI) might be an effective system for reducing pollutant emissions. After cold start, the SAI induces a post-oxidation of the **U**nburned **H**ydro**C**arbons (UHC) on the exhaust line accelerating the heat-up phase of the catalytic converter and thus reducing the time for reaching its light-off temperature. A modeling of SAI in the exhaust line post-combustion is necessary to optimize the shape of exhaust manifolds in order to get an efficient pollutant after treatment system and a minimum of UHC emissions. However, the modeling of such an air injection system is difficult since conditions for an efficient post-combustion inside the exhaust line are very restrictive. A pulsed gas flow in the exhaust manifold characterized by high turbulence and strong interactions between the exhaust streams from the different engine cylinders must be described. Additional to the mixing effects between the injected air and the exhaust gas, the chemical kinetics of the pollutants reacting with the injected air must be considered. Therefore, a mathematical description of UHC post-oxidation in the exhaust line demands the coupled modeling of the fluid dynamics and the chemical kinetics.

The fluid dynamics can be computed with a 3D turbulent combustion model coupled with a convenient description of the engine post-oxidation chemistry. This paper deals with the chemical kinetics part for which a reaction mechanism has been developed for the oxidation of typical pollutants (UHC, CO, $NO_x$) at thermodynamic conditions characteristic for engine post-oxidation. The exhaust gas composition is complex and might contain different UHC as well as combustion products and radicals. However, in order to minimize computation CPU times, not all the species present in the exhaust gases can be considered in a chemical reaction model. Therefore, n-heptane, and iso-octane as well as $NO_x$ have been chosen as key species representing the most important molecular groups of reactants at post-oxidation conditions. Other lighter hydrocarbons like methane or ethylene as well as CO and small size radicals (OH for example) which can also be present in the exhaust gases are included in the global mechanism.

## 2   Kinetic Model

The development of the two component kinetic model of hydrocarbon oxidation in the presence of $NO_x$ is described here. The mechanism for the oxidation of PRF (**P**rimary **R**eference **F**uels) mixtures (n-heptane and iso-octane) has been developed by Buda et al. [1] using EXGAS [3], an automatic mechanism generator. The computer aided generation of mechanisms, as well as the choice of the reaction types, the kinetics, and the a priori simplifications based on free radicals reactivity have been described extensively in the literature [3].

The kinetic model for the oxidation of PRF mixtures has been coupled with a $NO_x$ sub-mechanism. This mechanism was taken from a previous modeling work on the conversion of NO to $NO_2$ promoted by methane, ethane, ethylene, propane, and propene [4]. The kinetic model was primarily based on the GRI-MECH 3.0 [5] and the research performed by Dean and Bozzelli [6] and Atkinson et al. [7]. Coupling reactions between species involved in the alkane oxidation model and $NO_x$ were added. These reactions were derived from a mechanism published by Glaude et al. [2] for the oxidation of n-butane and n-pentane in presence of nitric compounds. Figure 1 shows schematically the coupling of the PRF oxidation mechanism with $NO_x$ species.

Mechanism for the oxidation of PRF mixtures.
Buda et al. [ 1 ]

Reactions coupling larger alcanes (C5, C7, C8) with + NO $_x$
presented in this paper deduced from Glaude [2]

$NO_x$ sub-mechanism
Dean [6] , Atkinson [7]

Mechanism for the oxidation of PRF mixtures in presence of NO $_x$

Figure 1: Coupling of the different reaction mechanisms

Figure 2 summarizes the coupling reactions with $NO_x$ species written:

- The reactions of alkylperoxy ROO· and hydroperoxy-alkylperoxy HOOQOO· radicals with NO leading to $NO_2$ and partially oxidized products. ROO· radicals react with NO forming $NO_2$ and alkoxy radicals RO·. HOOQOO· radicals reacting with NO are decomposed by a global reaction to $NO_2$, an hydroxyl radical OH, two formaldehyde molecules and the corresponding olefin. The rate constants chosen are identical to those proposed by Glaude for analogous reactions [2]. Alkoxy radicals RO· are decomposed by beta-scission with a rate constant proposed by Curran et al. [8]. The other reactions of alkoxy radicals considered by Glaude et al. [2] have been neglected.
- The reactions of resonance stabilized allylic radicals with $NO_2$ which yield NO, acrolein and an alkyl radical with the rate constant proposed by Glaude et al. [2].
- The conversion of aklyl radicals R· with $NO_2$ regenerating the alkoxy radical RO·. The rate constant was taken similar to those proposed by Glarborg et al. for the reaction $CH_3 + NO_2 = CH_3O + NO$ [9].

Thermochemical data was calculated using THERGAS [10]. In the case of nitrogen containing compounds, the data proposed by Marinov [11] have been used. The numerical calculations were performed using the CHEMKIN - software package [12].

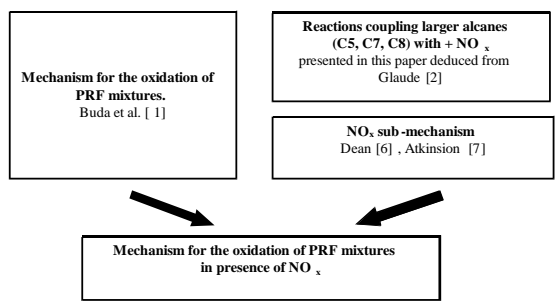

| | A (mol, cm, s, K) | b (-) | E (cal/mol) |
|---|---|---|---|
| Reaction base C0  -C2 with reactions of GRI3.0 included | | | |
| ·OOQOOH + NO  -> NO $_2$ + OH + 2  HCHO + olefin | 2.53e12 | 0.0 | -358.0 |
| ROO· + NO  -> RO ·+ NO | 2.53e12 | 0.0 | -358.0 |
| RO · -> aldehyde + alkyl | 2.0e13 | 0.0 | 15000 |
| Allylic radical + NO $_2$ -> alkyl + allylic aldehyde+ NO | 2.30e+13 | 0.0 | 0.0 |
| R·+NO $_2$ -> RO + NO | 4.0e13 | -0.2 | 0.0 |

Figure 2: Coupling of the different reaction mechanisms

## 3 Comparison between experimental and computed results

The mechanism has been tested against experimental jet stirred reactor [13] results for different pressures and temperatures. It has been validated for a temperature range from 700 K to 1100 K and pressures of 1 and 10 atm. The experimental data was obtained from Moréac [14]. The experimental setup has been described in detail in reference [13]. Figures 3a and 3b compare the calculated and the experimental concentration of n-heptane (Figure 3a) and CO (Figure 3b) for different initial NO concentrations. The results are presented for the stoichiometric mixtures of 1500 ppm of n-heptane with air in the presence of 0 ppm, 50 ppm and 500 ppm of NO as a function of temperature for a constant residence time of 0.2 s under atmospheric pressure.

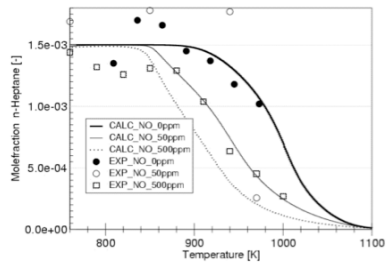

Figure 3a: Calculated (lines) and experimental (symbols) species concentration profiles of n-heptane in a jet-stirred reactor with equivalence ratio 1 at 1 atm with a residence time of 0.2 s in the presence of various amounts of NO: 0, 50 ppm and 500 ppm. Points refer to experimental data and lines to simulated data

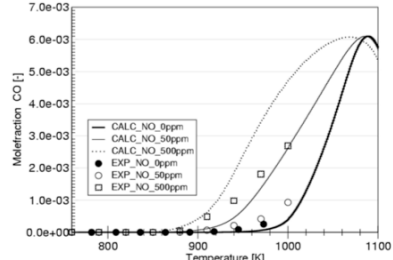

Figure 3b: Calculated (lines) and experimental (symbols) species concentration profiles of CO during the stoichiometric oxidation of n-heptane in a jet-stirred reactor at 1 atm with a residence time of 0.2 s in the presence of various amounts of NO: 0 ppm 50 ppm and 500 ppm. Points refer to experimental data and lines to simulated data

Experiments show that above 860 K, the oxidation of n-heptane is accelerated by the presence of NO, which can be observed by a reduced mole fraction of n-heptane and a simultaneous increased production of CO. As shown in Figures 3a and 3b, the kinetic model predicts well the accelerating effect of NO observed experimentally.

Figures 4a and 4b compare for different initial concentrations of NO the calculated and the experimental mole fractions of iso-octane and CO, respectively. The results are presented for the stoichiometric mixtures containing 1250 ppm of iso-octane with air in presence of 0 ppm and 50 ppm of NO as a function of temperature for a constant residence time of 0.2 s and 1 atm.

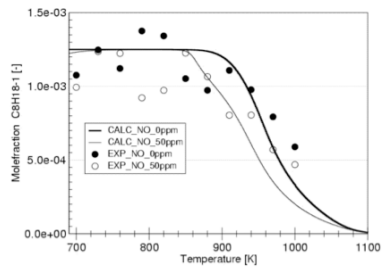

Figure 4a: Calculated (lines) and experimental (symbols) species concentration profiles of iso-octane in a jet-stirred reactor with equivalence ratio 1 at 1 atm with a residence time of 0.2 s in the presence of various amounts of NO: 0 ppm and 50 ppm. Points refer to experimental data and lines to simulated data

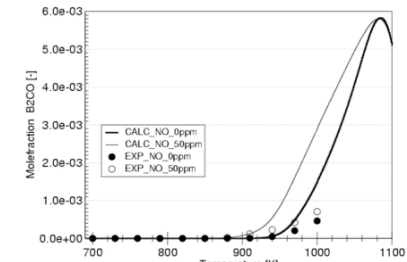

Figure 4b: Calculated (lines) and experimental (symbols) species concentration profiles of CO during the stoichiometric oxidation of iso-octane in a jet-stirred reactor at 1 atm with a residence time of 0.2 s in the presence of various amounts of NO: 0 ppm and 50 ppm. Points refer to experimental data and lines to simulated data

As in the case of n-heptane, above 880 K, the oxidation of iso-octane is accelerated by the presence of NO. As shown in Figures 4a and 4b, the simulations clearly predict the catalyzing effect of NO on the oxidation of iso-octane which for increasing amounts of NO results in declined iso-octane mole fraction and simultaneous increased CO formation.

The kinetic model has also been validated against HCCI-engine experiments of Dubreuil et al. [15] for a fuel mixture of n-heptane (75%vol) and iso-octane (25%vol) and for an equivalence ratio of 0.3. The initial NO concentration was varied from 0 ppm up to 500 ppm. Figure 5 compares the calculated and the experimental ignition delays for the cool and main flame as a function of the initial NO concentration. The ignition delays are presented in Crank-Angle-Degrees-after-Bottom-Dead-Center (CAD ABDC).

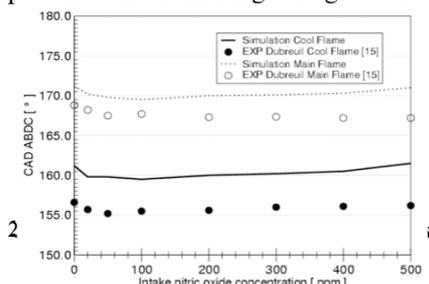

Figure 5: Comparison between modeling and experimental data obtained for different NO concentrations for a n-heptane/iso-octane fuel mixture. Points refer to experimental data and lines to simulated data

For small NO concentrations NO (20 ppm and 50 ppm), a decrease of the cool and main flame ignition delay can be observed. For NO concentrations from 100 ppm up to 500 ppm, only little influence of NO on the ignition delay is noticeable. Compared to experiments, the model predicts an off-set of the ignition delays of about 5 CAD ABCD. The crank angle delay between cool and main flames as well as the general trends of NO-influence are in good agreement with the experimental data.

## 4   Conclusions and future work

The mechanism presented in this paper reproduces well the NO effect on the kinetics of n-heptane and iso-octane known from experiments performed in a Jet Stirred Reactor. The simulations performed for a HCCI-engine for a mixture of n-heptane and iso-octane were also in good agreement with experimental results. Further validations on mixtures of n-heptane / iso-octane will be performed. A model extension towards ternary mixtures of n-heptane / iso-octane / toluene in presence of NO is in progress. Once the chemical reaction mechanism is validated, thermodynamic tables will be deduced and coupled to a 3D-CFD-code. The tables generated will be used as a database for a turbulent combustion model implemented in an engine CFD-code. The ensemble of generated tables and CFD-code will be used as a tool for understanding physical and chemical processes of post-oxidation in IC-engines.